\documentclass{jps-cp}
\usepackage{txfonts}

\title{On Schottky Noise and Shot Noise}

\author{Xiangcheng \textsc{Chen}}

\inst{Institute of Modern Physics, Chinese Academy of Sciences, Lanzhou 730000, China}

\email{cxc@impcas.ac.cn}

\abst{
  Schottky noise is a common term widely acknowledged by the community of accelerators, especially of circular machines.
  It is referred to the incoherent signal arising from the circulating charged particles in the accelerator.
  The noise is named after W.~Schottky, who first discovered a then new type of noises in direct electric current and called it shot noise himself.
  Because of this, the Schottky noise is considered as an alias for the shot noise in accelerator literatures.
  But is it really so?
  In this essay I compare the two noises side-by-side by formulas and argue that even though they arise from the same discrete nature of charge carriers, the noise spectral patterns are inherently distinct, whereby the synonymity of these two terms is questionable.
}

\kword{Schottky noise, shot noise, power spectral density}

\begin{document}
\maketitle

\section{Introduction}

In the seminal paper published in 1918 by the German physicist Walter Schottky, he reported the discovery of a current fluctuation in a vacuum tube with a thermal cathode emitting electrons and an anode collecting them shortly after~\cite{Schottky-1918}.
Schottky soon realized that such a fluctuation is purely statistical, which can be attributed to the quantal nature of each electron being randomly transmitted from the cathode to the anode.
On the macroscopic scale the cathode ray forms a unidirectional and constant electric current, whereas on the microscopic scale the electrons may arrive at the anode in a different bulk size from time to time, resulting in a small variation riding on the average current.
Schottky analyzed this phenomenon and concluded that the noise is absolutely white with a mean power proportional to the average current itself, instead to the squared current as normally dictated by the Ohm's law.

Several decades later when physicists wanted to build a synchrotron to accelerate charged particles, they would, by analogy, expect a similar noise pattern present in the beam current since the charge carriers in that case are also discrete.
Indeed in 1971, physicists had observed the predicted noise in the ISR (intersecting storage ring) at CERN~\cite{Caspers-2012}.
Since then, the community of accelerator physics became aware of this type of noise ubiquitous in circular machines, and called it Schottky noise in name of its discover decades ago.
However, quite a number of authors publishing in this field deceptively state, in a way, that the Schottky noise is actually the shot noise, which should be disputed in my opinion.
In the following, I shall first point out the mathematical formalism of these two noises in the statistical language, then reveal their distinctions in terms of noise spectral density, and finally conclude with personal remarks.

\section{Shot Noise}

The apparatus used in Schottky's experiment can abstractly be modeled as a detector receiving a bunch of incoming identical point charges $q$ in a stochastic fashion.
Let $N$ denote the number of point charges in an arbitrarily given temporal interval $T$.
As $T$ extends infinitely long, the \emph{macroscopic} current on average is assumed to tend to a constant value $\bar{I}$:
\begin{equation}
  \bar{I} = \lim_{T\to\infty}\frac{Nq}{T}.
\end{equation}
But looking closer, the arrival time $t_i$ of each point charge $i$ is randomly distributed.
As a consequence, the \emph{microscopic} current $I(t)$ as a function of time $t$ is the summation of infinite pulses:
\begin{equation}
  I(t) = \sum_{i=1}^{N\to\infty} q\delta(t-t_i),
\end{equation}
where $\delta(t)$ is the Dirac function.

The \emph{Wiener-Khinchin theorem} asserts that the power spectral density $S_o(f)$ of $I(t)$ is the Fourier transform of its autocorrelation function $R_o(\tau)$~\cite{Papoulis-2002}.
By definition, $R_o(\tau)$ can be calculated as
\begin{equation}
  R_o(\tau) = \operatorname{E}\left\{ I(t+\tau)I(t) \right\} = \sum_{i,j=1}^{N}q^2\operatorname{E}\left\{ \delta(t+\tau-t_i)\delta(t-t_j) \right\}.
\end{equation}
If additionally the ergodicity is presumed, which is almost always fulfilled in reality, the above expectation can be replaced by the time average:
\begin{equation}
  R_o(\tau) = \lim_{T\to\infty}\frac{q^2}{T}\sum_{i,j=1}^{N}\int_{-T/2}^{T/2}\delta(t+\tau-t_i)\delta(t-t_j)dt
  = \lim_{T\to\infty}\frac{q^2}{T}\sum_{i,j=1}^{N}\delta(\tau-t_i+t_j)
\end{equation}
By virtue of $R_o(\tau)$, $S_o(f)$ can be written as
\begin{equation}
  \begin{split}
    S_o(f) &= \int_{-\infty}^{\infty}R_o(\tau)e^{-i2\pi f\tau}d\tau \\
    &= \lim_{T\to\infty}\frac{q^2}{T}\sum_{i=j=1}^{N}\int_{-\infty}^{\infty}\delta(\tau)e^{-i2\pi f\tau}d\tau
    + \lim_{T\to\infty}\frac{q^2}{T}\sum_{i\neq j}\int_{-\infty}^{\infty}\delta(\tau-t_i+t_j)e^{-i2\pi f\tau}d\tau.
  \end{split}
\end{equation}
The first integral on the right side can easily be reduced to
\begin{equation}
  \lim_{T\to\infty}\frac{q^2}{T}\sum_{i=j=1}^{N}\int_{-\infty}^{\infty}\delta(\tau)e^{-i2\pi f\tau}d\tau = \lim_{T\to\infty}\frac{Nq^2}{T} = q\bar{I}.
\end{equation}
For the second integral, it is in fact the summation of infinite, random phases of the form $e^{-i2\pi f(t_i-t_j)}$, which will effectively vanish as $T\to\infty$.

In the end, the noise spectral density simply reads
\begin{equation}
  S_o(f) = 2q\bar{I},
  \label{eq:psd-shot}
\end{equation}
where the factor $2$ arises from folding the negative half of the spectrum to the positive part, since negative frequencies is not a observable in practice.

\section{Schottky Noise}

One of the most prominent features of the beam current in a circular machine, like a storage ring, is that every stored point charge $q$ circulates repetitively.
For the sake of simplicity, a coasting beam consisting of $N$ point charges is presumed to be revolving at a defined frequency $f_r$ in the ring.
A Schottky detector is installed at a fixed location to register the transit of every point charge.
Consequently, the microscopic beam current $I(t)$ can be expressed as
\begin{equation}
  I(t) = \sum_{i=1}^{N}\sum_{n=-\infty}^{\infty}q\delta(t-t_i-nT),
\end{equation}
where the index $i$ enumerates point charges, and $n$ denotes every repetition of transit by the same point charge at the period $T=1/f_r$.
By the assumption of coasting beam, $t_i$ is expected to be uniformly distributed within the interval of $(0, T)$.

The autocorrelation function $R_c(\tau)$ of $I(t)$ is similarly given as
\begin{equation}
  R_c(\tau) = \operatorname{E}\left\{ I(t+\tau)I(t) \right\} = \sum_{i,j=1}^{N}\sum_{n,m=-\infty}^{\infty}q^2\operatorname{E}\left\{ \delta(t+\tau-t_i-nT)\delta(t-t_j-mT) \right\}.
\end{equation}
The expectation value of the second equality in the above shall be calculated in two parts.
First consider the noise power contributed by the same point charge, i.e.\ $i=j$:
\begin{equation}
  \begin{split}
    \operatorname{E}\left\{ \delta(t+\tau-t_i-nT)\delta(t-t_i-mT) \right\} &= \frac{1}{T}\int_{0}^{T}\delta(t+\tau-t_i-nT)\delta(t-t_i-mT)dt_i \\
    &= \begin{cases}
      \frac{\delta(\tau-nT+mT)}{T} & m=\lfloor\frac{t}{T}\rfloor \\
      0 & \mathrm{otherwise}
    \end{cases},
  \end{split}
\end{equation}
where $\lfloor x\rfloor$ denotes the floor function.
Then consider the noise power of the cross term, i.e.\ $i\neq j$:
\begin{equation}
  \begin{split}
    \operatorname{E}\left\{ \delta(t+\tau-t_i-nT)\delta(t-t_j-mT) \right\} &= \frac{1}{T^2}\int_{0}^{T}\int_{0}^{T}\delta(t+\tau-t_i-nT)\delta(t-t_j-mT)dt_idt_j \\
    &= \begin{cases}
      \frac{1}{T^2} & n=\lfloor\frac{t+\tau}{T}\rfloor\textrm{ and }m=\lfloor\frac{t}{T}\rfloor \\
      0 & \mathrm{otherwise}
    \end{cases}.
  \end{split}
\end{equation}
Aggregating these two parts gives rise to
\begin{equation}
  R_c(\tau) = \sum_{i=1}^{N}\sum_{n=-\infty}^{\infty}\frac{q^2}{T}\delta(\tau-nT) + \sum_{i\neq j}\frac{q^2}{T^2} = \frac{Nq^2}{T}\sum_{n=-\infty}^{\infty}\delta(\tau-nT) + 2!\binom{N}{2}\frac{q^2}{T^2}.
\end{equation}
The second term being constant on the right side of the above equation will, after Fourier transform, lead to a $\delta$-spike at zero frequency in the noise spectrum.
Hence it belongs to the DC (direct current) component of the beam current.
In contrast, the first term represents the Schottky fluctuation on top of the DC current, whereby is of concern in this essay.

The power spectral density $S_c(f)$ of the Schottky noise is simply
\begin{equation}
  S_c(f) = \frac{Nq^2}{T}\int_{-\infty}^{\infty}\sum_{n=-\infty}^{\infty}\delta(\tau-nT)e^{-i2\pi f\tau}d\tau = \frac{Nq^2}{T^2}\sum_{k=-\infty}^{\infty}\delta(f-kf_r) \simeq 2q\bar{I}f_r\sum_{k=1}^{\infty}\delta(f-kf_r),
  \label{eq:psd-schottky}
\end{equation}
where in the last step, the DC component is dropped out, the negative frequencies are superposed to the corresponding positive ones, and the macroscopic current $\bar{I}=Nq/T$ is substituted in.

\section{Discussion}

Following the above derivations, one finds that the shot noise and the Schottky noise share the same origin in the sense of discreteness of the charge carriers.
However, by comparison between the power spectral density of the shot noise in Eq.~(\ref{eq:psd-shot}) and the Schottky noise in Eq.~(\ref{eq:psd-schottky}), it is also evident that these two types of noises are quite distinct in the frequency domain.
The shot noise manifests a plain, frequency-independent spectrum, which is an important character of the white noise.
On the contrary, the Schottky noise spectrum consists of an infinite number of spikes, denoted by its harmonics, equally distanced by the revolution frequency.
According to Eq.~(\ref{eq:psd-schottky}), the integral power over any frequency band of width $f_r$, which just covers one of those spikes, is $2q\bar{I}f_r$---a value matches the one for the shot noise by means of Eq.~(\ref{eq:psd-shot}).
This can be loosely understood as due to the arrival periodicity of the charge carriers at the detector in the case of Schottky noise, the corresponding power spectral density also demonstrates some sort of periodicity by condensing the evenly distributed power into very sharp spikes.
When taking into account a tiny momentum spread, those spikes smear out into bands whose width increases proportionally to the harmonic while the height inversely proportionally decreases so as to preserve the integral power in any band~\cite{Chattopadhyay-1985}.
Consequently the Schottky noise contains vast information about the beam parameters, in which the non-invasive Schottky diagnostics for machine tuning~\cite{Caspers-2008} and Schottky spectroscopy for physics experiments~\cite{Bosch-2003} are deeply rooted.
In this sense, the \emph{Schottky signal} seems to be a more proper name for such a fluctuation, and better stresses the intrinsic difference from the shot noise.

\section{Conclusion}

To summarize, the Schottky noise is not the same as, but a special realization of, the shot noise.


\begin{thebibliography}{9}
		\bibitem{Schottky-1918} W.~Schottky, Ann.\ Phys.\ (Berlin) \textbf{362}, 541 (1918).
		\bibitem{Caspers-2012} F.~Caspers and D.~M\"ohl, Eur.\ Phys.\ J.\ H \textbf{36}, 601 (2012).
    \bibitem{Papoulis-2002} A.~Papoulis and S.~U.~Pillai, Probability, Random Variables and Stochastic Processes, 4th edition (McGraw-Hill, New York, 2002).
    \bibitem{Chattopadhyay-1985} S.~Chattopadhyay, AIP Conf.\ Proc.\ \textbf{127}, 467 (1985).
		\bibitem{Caspers-2008} F.~Caspers, Proc.\ CAS '08: Beam Diagnostics, 407 (2008).
		\bibitem{Bosch-2003} F.~Bosch, J.\ Phys.\ B \textbf{36}, 585 (2003).
\end{thebibliography}
\end{document}